\documentclass[draft]{agujournal2019}
\usepackage{url} 
\usepackage{amsmath,amssymb}
\drafttrue

\journalname{Geophysical Research Letters}

\begin{document}

%
%


\title{Ranking IPCC Model Performance Using the Wasserstein Distance}

%
%




\authors{G. Vissio\affil{1,2}, V. Lembo\affil{1,3}, V. Lucarini\affil{1,4,5}, and M. Ghil\affil{6,7}}

\affiliation{1}{CEN, Meteorological Institute, University of Hamburg, Hamburg, Germany}
\affiliation{2}{{\color{black}Institute of Geosciences and Earth Resources (IGG) - National Research Council (CNR), Torino, Italy}}
\affiliation{3}{{\color{black}Institute of Atmospheric Sciences and Climate (ISAC) - National Research Council (CNR), Bologna, Italy}}
\affiliation{4}{Department of Mathematics and Statistics, University of Reading, Reading, UK}
\affiliation{5}{Centre for the Mathematics of Planet Earth, University of Reading, Reading, UK}
\affiliation{6}{Geosciences Department and Laboratoire de M\'et\'eorologie Dynamique (CNRS and IPSL), Ecole Normale Sup\'erieure and PSL University, Paris, France}
\affiliation{7}{Department of Atmospheric \& Oceanic Sciences, University of California at Los Angeles, Los Angeles, USA}





\correspondingauthor{{\color{black}Valerio Lucarini}}{{\color{black}v.lucarini@reading.ac.uk}}




\begin{keypoints}
\item New method for evaluating the skill of a climate model;  
\item Climate model ranking according to the selected variables of interest;
\item Ability to highlight model deficiencies through emphasis on specific geographical regions and climatic variables.
\end{keypoints}

%
%


\begin{abstract}
We propose a methodology for intercomparing climate models and evaluating their performance against benchmarks based on the use of the Wasserstein distance (WD). This distance provides a rigorous way to measure quantitatively the difference between two probability distributions. The  proposed approach is flexible and can be applied in any number of dimensions; it allows one to rank climate models taking into account all the moments of the distributions. By selecting  the combination of climatic variables and the regions of interest, it is possible to highlight specific model deficiencies. The WD  enables a comprehensive evaluation of climate model skill. We apply this approach to a selected number of physical fields, ranking the models in terms of their performance in simulating them, and pinpointing their weaknesses in the simulation of some of the selected physical fields in specific areas of the Earth.
\end{abstract}

%
%

%


%
%
%
%

\section{Introduction and motivation}\label{sec:intro}

Advanced climate models differ in the choice of prognostic equations and in the methods for their numerical 
solution, in the number of processes that are parametrized and the choice of the physical parametrizations, as well as in the way the models are initialized, to mention just their most important aspects. Comparing the performance of such models is still a major challenge for the climate modeling community \cite{Held2005}. 

Model inadequacies, which may lead to large uncertainties in the model's predictions, result from structural errors --- certain processes are incorrectly represented or not represented at all --- as well as from parametric uncertainties, i.e., the use of incorrect values for the parameters associated with processes that are correctly formulated in the model \cite{Lucarini2013, Ghil.Luc.2020}. Intercomparing climate models and auditing them individually is essential for understanding which ones are more skillful in answering the specific climate question under study.

{\color{black}The need for testing systematically model performance 
has led the community to join forces through} the Coupled Model Intercomparison Project (CMIP), which is currently in its sixth phase \cite{Eyring2016a}. Dozens of modeling groups have agreed on a concerted effort to provide numerical simulations with standardized experimental protocols representative of specified climate forcing scenarios.

{\color{black}There is no standard suite of metrics to evaluate
climate model performance nor, a fortiori, to decide whether a model does have skill in predicting future climate change. \citeA{Lucarini2007} suggested that testing a climate model's performance requires considering a mixture of global and process-oriented metrics. \citeA{Gleckler2008} proposed a multidimensional metric based on the comparison of the spatio-temporal variability of many climatic fields with respect to reference datasets, and found that creating a scalar comprehensive metric is nontrivial.  
\citeA{Eyring2016b} and \citeA{Eyring2020} have combined metrics and diagnostic tools designed to assess specific features of the climate system, whilst \citeA{Lembo2019} have provided a tool  to test the  models' skill in representing the thermodynamics of the climate system. 

Hence, }
it seems highly desirable  
to have a scalar metric that summarizes the  information associated with model performance and that satisfies the mathematical axioms associated with the concept, as for the usual Euclidean distance. These axioms are listed in Text S1 of the Supplementary Information (SI) and they are satisfied by the root-mean-square distance, known as an $L_2$ metric in mathematics. The latter distance, though, is not appropriate for describing fully the difference between two distribution functions, while other metrics used in the climate sciences are not genuine distances, i.e., they do not satisfy the axioms above.

We propose a new metric to assess a climate model's skill by taking into account every moment of a distribution and measuring 
the gap between it and another distribution of reference.  
The two distributions will be chosen here to describe model features, on the one hand, and the "real world," on the other, 
with the latter distribution being based on  raw observations and/or a reanalysis thereof.

%

\citeA{Ghil2015} originally proposed the idea of using the Wasserstein distance (hereafter WD) \cite{Dobrushin.1970,Kantorovich.1942, Villani2009} in the context of the climate sciences as a way to generalize the traditional concept of equilibrium climate sensitivity \cite{Ghil.Luc.2020} in the presence of a time-dependent forcing, such as seasonal or anthropogenic forcing. \citeA{Robin2017} used the WD to compute the difference between the snapshot attractors of the \citeA{Lorenz.1984} model for different time-dependent forcings, providing a link between nonautonomous dynamical systems theory and optimal transport. \citeA{Vissio2018a} used the WD to evaluate the skill of a stochastic parametrization for a fast-slow system.  {\color{black}\citeA{Ning.ea.2014} proposed the use of the WD to quantify  model error in variational data assimilation and presented an insightful application in the case of advection-diffusion dynamics with systematic errors in the velocity and diffusivity parameters.} Please see Text S1 in the SI for further background on the WD.

  {\color{black}Well-known WD drawbacks are (a) its computational requirements, which increase dramatically with the number of points used to construct the empirical distributions; and (b) the curse of dimensionality: the amount of data needed to explore accurately a higher dimensional phase space grows exponentially with the number of dimensions. Concerning (a), \citeA{Vissio2018a} and \citeA{Vissio2018b} have shown that the computational requirements are greatly reduced through data binning on a grid. 
  As for (b), the WD will be calculated in a reduced phase space defined by the physical variables we wish to take into account in the evaluation of the model. The possibility of freely choosing the variables of interest makes the WD  a flexible candidate for evaluating a climate model's skill.} 
  

The WD-based metric can complement the existing methods used for intercomparing climate models, {\color{black} such as ranking of model performances with respect to the root-mean-square-error of the median of an ensemble \cite{Flato2013} or weighted ensemble averaging schemes based on models' discrepancy from observations \cite{Knutti2017}.}
This letter is structured as follows. Data are presented in Sec. \ref{Dat}, methods in Sec. \ref{WasDis}, results in Sec. \ref{Rank}, and conclusions in Sec. \ref{sec:conclude}. The Supporting Information (SI) provides technical details.


\section{Data} \label{Dat}

The WD methodology is presented in Sec.~\ref{WasDis}. It is applied here to three climate fields:
\begin{itemize}
\item Near-surface air temperature; 
\item Precipitation; and
\item Sea ice cover, computed from the sea ice area fraction. 
\end{itemize}
The corresponding daily mean fields are available in the CMIP5 simulations for historical and RCP85 forcings \cite{Taylor2012} and they are ranked with respect to the distance from reference daily datasets, specifically 
European Centre for Medium-Range Weather Forecasts Re-Analysis (ERA) Interim for the temperature \cite{Dee2011}; Global Precipitation Climatology Project (GPCP) for the precipitation \cite{Adler2003}; and Ocean and Sea Ice - Satellite Application Facility (OSI-SAF) for the sea ice cover \cite{EUMETSAT2017}. In order to further support the comparison and provide a benchmark, we analyzed the WD with respect to the National Center of Environmental Prediction (NCEP) Reanalysis 2 \cite{Kanamitsu2002}. 

The fields are averaged on four distinct domains: (i) Global; (ii) Region between $30$~S and $30$~N (Tropics);
(iii) Region between $30$~N and $90$~N (Northern extratropics); and (iv) Arctic -- used only for sea ice extent.
While temperature and precipitation analyses involve 30 models, taking into account sea ice extent allows to analyze just 22 models, due data availability.%
datasets.
The time range spans 18 years, from 01/01/1997 to 12/31/2014. After the spatial averaging, the model datasets are obtained by concatenating the historical runs, from 1997 to 2005, and the RCP85 runs, from 2006 to 2014. The acronyms of the models considered here 
are given in Table~S1 of Text S2 in the SI. 

The samples used in the WD calculations are drawn by performing a \citeA{Ulam1964} discretization of the phase space involved in each separate test. To do so, a regular grid is superposed over all the datasets used in the test and its upper and lower limits, respectively, are fixed slightly above and below the maximum and minimum values among all the datasets used in it. Each dimension of the grid is then equally divided into 20 intervals; this yields $20^m$ $m$-dimensional cubes, where $m$ is the number of fields taken into account in the test. 
These $20^m$ hypercubes provide the sample for each test. The results we present here are weakly sensitive to the specifics of the gridding. Nonetheless, a too coarse gridding removes a lot of the information we want to retain and analyse; a too fine gridding, instead, increases substantially the computing requirements, without making much statistical sense.

In order to highlight the flexibility and reliability of the method, we are going to calculate the WD distances in one-, two- and three-dimensional phase space, and work with different field combinations averaged over distinct areas of the Earth.

\section{Wasserstein distance} \label{WasDis}

Our objective is to create a ranking of the CMIP5 IPCC models based on their skill to reproduce the statistical properties of selected physical quantities. The reference distribution for these quantities is given by reanalysis and observational datasets, as explained in Sec.~\ref{Dat}; their WD to these datasets 
is a measure of the models' ability to reproduce these reference distributions. One can also describe this distance as the minimum "effort" to morph one distribution into the other \cite{Monge1781}. We present below a very simplified account of the theory.

The optimal transport cost \cite{Villani2009} is defined as the minimum cost to move the set of {\color{black}$n$} points from one distribution to another into an {\color{black}$m$}-dimensional phase space. In the case of two discrete distributions, we write their measures $\mu$ and $\nu$ as
\begin{equation} 
\mu =\sum\limits_{i=1}^n \mu_i \delta_{x_i} , \qquad
\nu =\sum\limits_{i=1}^n \nu_i \delta_{y_i} ;
\end{equation}
here $\delta_{x_i}$ and $\delta_{y_i}$ are Dirac measures associated with a pair of points $(x_i, y_i)$, whose  fractional mass is $(\mu_i, \nu_i)$, respectively, and $\sum_{i=1}^n \mu_i=\sum_{j=1}^n \nu_j=1${\color{black}, where all the terms in the sum are nonnegative.} 
Using the definition of Euclidean distance
\begin{equation} 
d(\mu,\nu)=\left[ \sum\limits_{i=1}^n (x_i-y_i)^2 \right] ^ {\frac{1}{2}},
\end{equation}
we can write down the quadratic WD for discrete distributions:
\begin{equation} 
W_2(\mu,\nu)= \left\lbrace \inf_{\gamma_{ij}} \sum\limits_{i,j} \gamma_{ij} [d(x_i,y_j)]^2 \right\rbrace ^ {\frac{1}{2}}. \label{wd2discrete}
\end{equation}

Here {\color{black}$\gamma_{ij}$, {\color{black}is a \textit{transport protocol}, which defines how the fraction of mass is transported from $x_i$ to $y_j$}, while $d(x_i,y_j)$ is the Euclidean distance between a single pair of locations. The transport protocol realizing the minimum in Eq.~\eqref{wd2discrete} is called the \textit{optimal coupling}; see a visual explanation in Fig.~S1 of the SI.} 

We perform the Ulam discretization described above 
that allows us to shift from the distance between different distributions of points {\color{black}given by the time series to the distance between measures that can be estimated from such distributions (via data binning), while sticking to a discrete optimization problem, as discussed below. See \citeA{Santambrogio2015} for a survey of numerical methods for computing the WD.} We thus proceed to quantify to what extent the measure of the observations and reanalysis from Sec.~\ref{Dat}, projected on the variables of interest, differs from the corresponding measures for the climate models. 

The estimate of the coarse-grained probability of being in a specific grid box is given by the time fraction spent in that  box \cite{Ott1993,Strogatz2015}. In fact, the WD does provide robust results even with a very coarse grid \cite{Vissio2018a,Vissio2018b}. Therefore, in the case at hand, the locations $x_i$ and $y_j$ will indicate the cubes' centroids, while $\gamma_{ij}$ indicate the corresponding densities of points. {\color{black}If $(k_1,k_2,\ldots,k_m)$ are indices running from $1$ to $n$, the cube with position $x$ in the $m$-dimensional space will be identified by the $m-$tuple $k^x_1,k^x_2,\ldots,k^x_m$. We then define $d(x,y)^2=\sum_{l=1}^m(k^x_l-k^y_l)^2$.} To further simplify the computations, we exclude all the grid boxes containing no points at all. Finally, we divide the distance by $n$; 
therefore, the one-, two- and three-dimensional WDs take values between a minimum of $0$ and a maximum equal to $1$, $\sqrt{2}$ and $\sqrt{3}$, respectively.

We used a suitably modified version of the Matlab software written by G. Peyr\'e --- {\color{black}available at  \url{https://nbviewer.jupyter.org/github/gpeyre/numerical-tours/blob/master/matlab/optimaltransp_1_linprog.ipynb}} --- to perform the calculations. The modifications include the data binning and the estimation of the measures, as well as adapting to 
a dimension $m \ge 2$. 


\section{Ranking the models} \label{Rank}

\begin{figure}
\noindent\includegraphics[width=\textwidth]{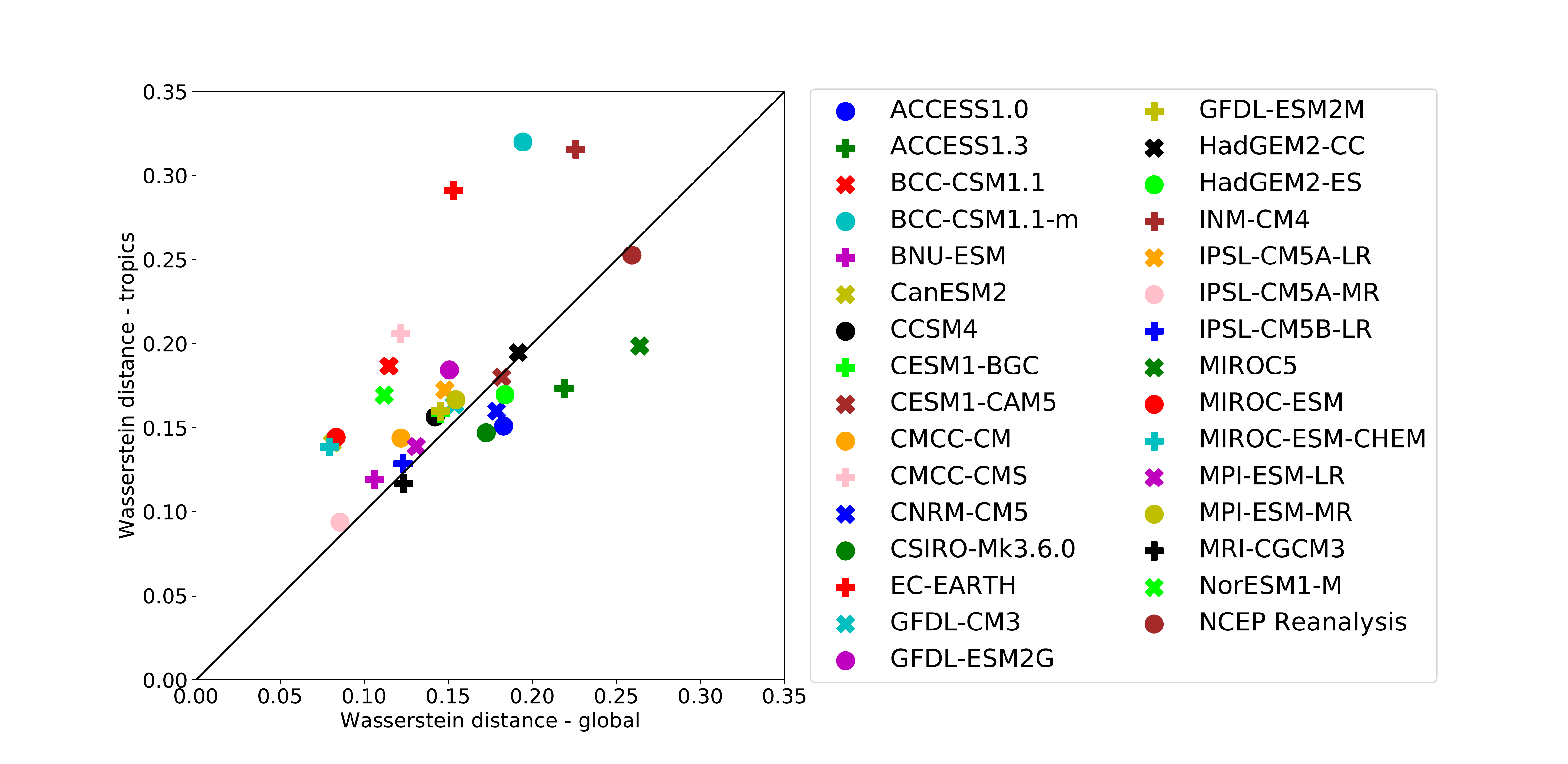}
\caption{Two-dimensional Wasserstein distance (WD) for the temperature and precipitation fields, 
      averaged over the globe (horizontal axis) and over the Tropics (vertical axis). 
      The acronyms of the models 
      used are spelled out in Text S2 of the SI.}
\label{Figure_1}
\end{figure}

Figure~\ref{Figure_1} shows the WD calculated in the two-dimensional phase space composed by the temperature and precipitation fields, averaged over the whole Earth and the Tropics, for each CMIP5 model. In order to provide a benchmark, we chose to include the WD results between the NCEP reanalysis and the references given by the {\color{black}ERA temperature and GPCP precipitation fields}, respectively. 

 Somewhat surprisingly, the NCEP reanalysis yields the largest values in both distances. Thus, the average CMIP5 distance to the 
 combined ERA-and-GPCP reference datasets is $0.149$, while the NCEP distance is $0.259$, exceeded only by the value $0.264$ given by the MIROC5 model; see Table~S1 in the SI for the list of models. 
 Note that the one-dimensional WDs of the NCEP Reanalysis for the globally averaged temperature and precipitation equal 0.033 and 0.255, respectively, which indicates the inadequacy of the NCEP dataset in representing the statistics of precipitation. {\color{black}Despite the well-known difficulties with simulating the very rough precipitation field by using the still fairly coarse CMIP5 models \cite{Neelin.2013,Mehran.2014}, the results point to the overall accuracy reached by CMIP5 simulations when dealing with global averages of temperatures and precipitation. }
 
 
We evaluate next the problems still encountered by CMIP5 models in reproducing key aspects of tropical dynamics \cite{Tian2020}. Averaging the data over the Tropics, we obtain the ranking on the vertical axis in Fig.~\ref{Figure_1}. {\color{black} The WD distance is for most datasets larger than when looking at globally averaged quantities (the models' mean is $0.173$), and underline the poorer CMIP5 model performances in this region. With few exceptions, the models seem less reliable 
in the Tropics, where three of the models exceed the NCEP Reanalysis distance. This distance is very similar to what has been found for the globally averaged case.} 

\begin{figure}
\noindent\includegraphics[width=\textwidth]{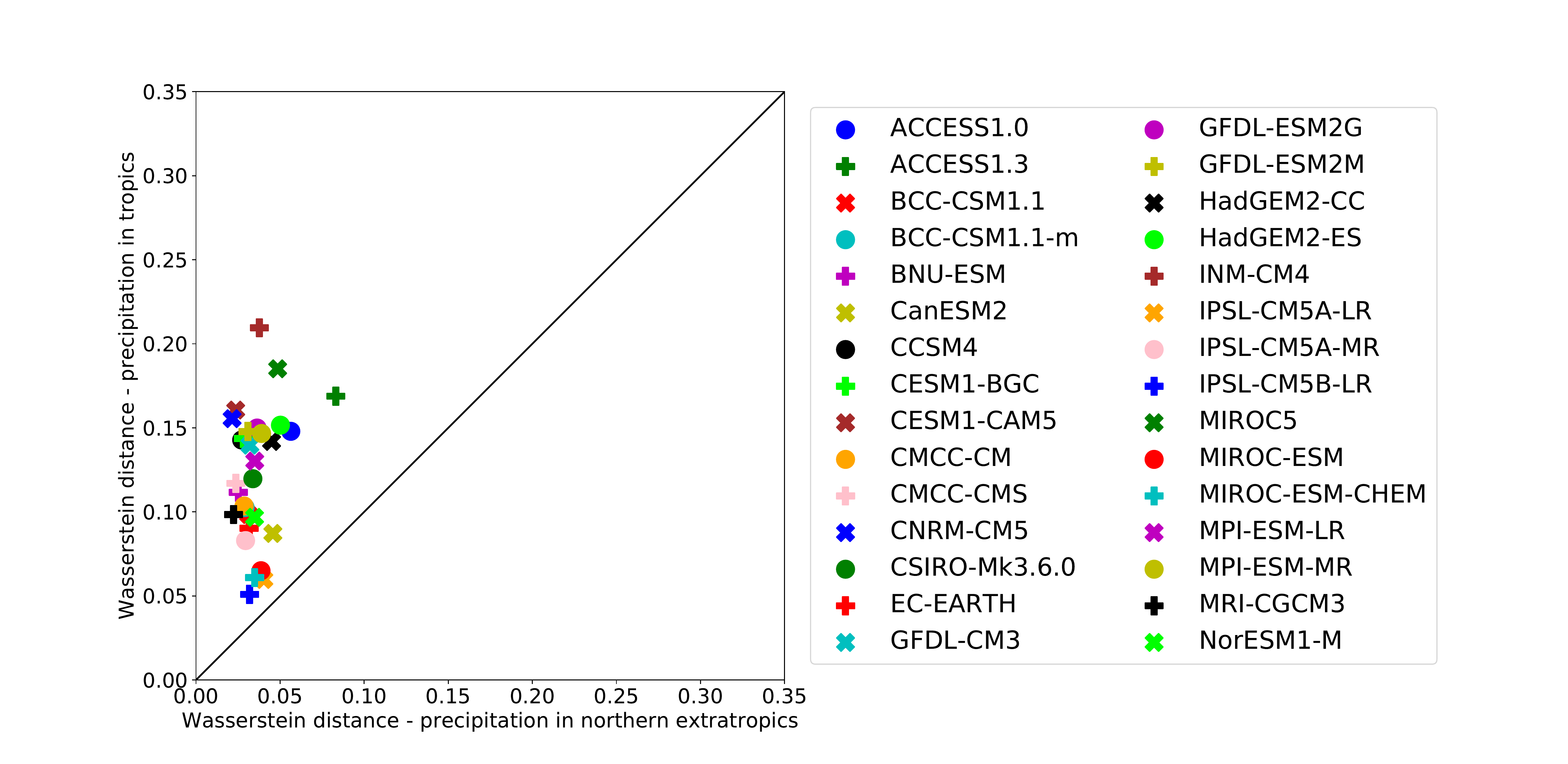}
\caption{One-dimensional WD for precipitation averaged over the Northern extratropics 
      (from $30$~N to $90$~N) on the horizontal axis and over 
      the Tropics (from $30$~S to $30$~N on the vertical axis).}
\label{Figure_2}
\end{figure}

\begin{figure}
\noindent\includegraphics[width=\textwidth]{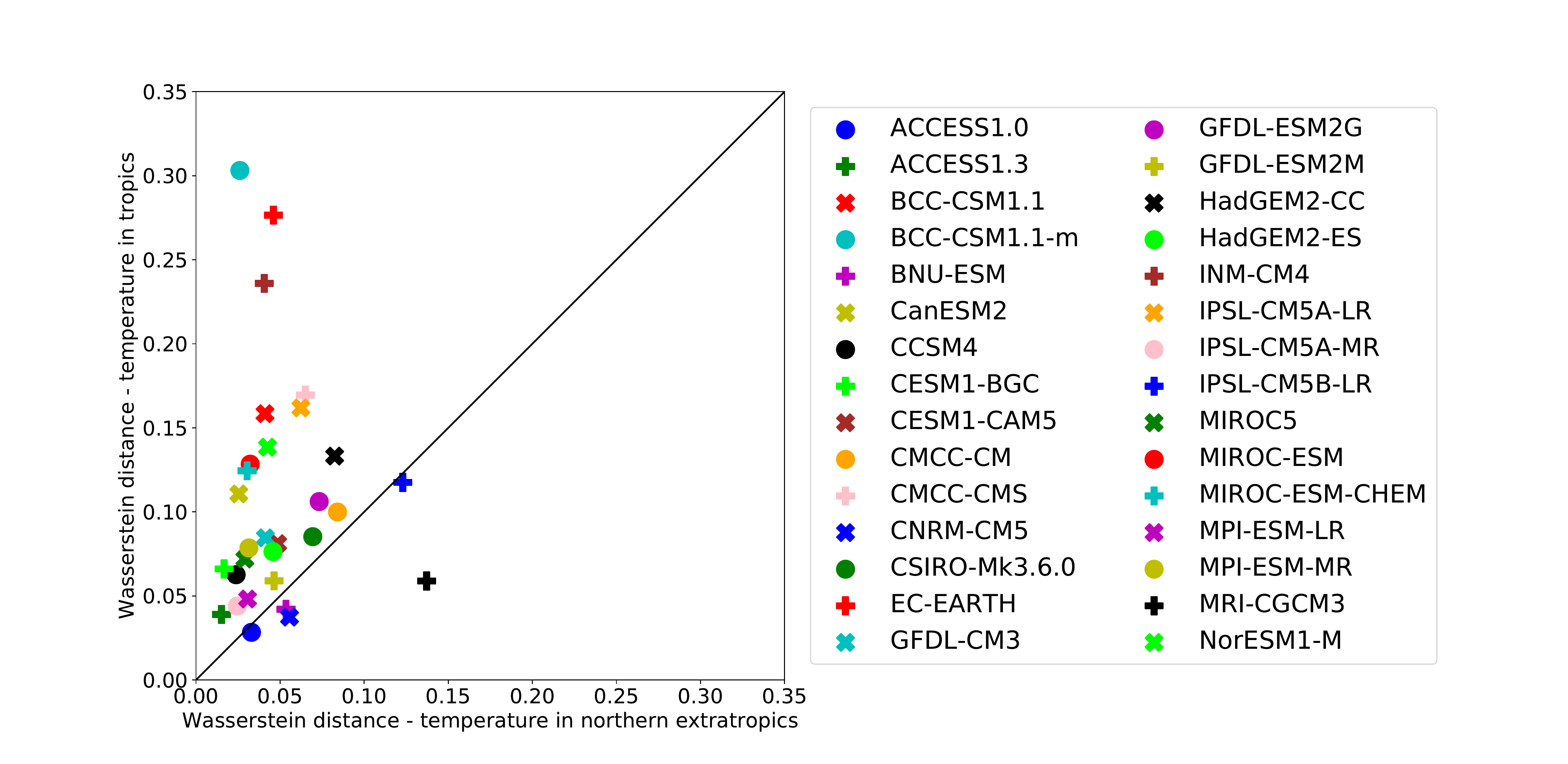}
\caption{
      Same as Fig.~\ref{Figure_2} but for the temperature field.}
\label{Figure_3}
\end{figure}

{\color{black} Next we show how the WD can be used to perform comparative analyses of the performance of a given model or of a group of models with respect to different climatic observables.} Focusing on the relative performance of temperature and precipitation in the Tropics vs the Northern Hemisphere extratropics, Figs.~\ref{Figure_2} and \ref{Figure_3} illustrate one-dimensional WDs computed in the former vs the latter region. Using the diagonal line indicating equal values for the two distances as a reference, we can easily check in Fig.~\ref{Figure_2} that, for all CMIP5 models, the precipitation field is less well reproduced in the Tropics than in the extratropics: it is extremely challenging to reproduce accurately the statistics of by-and-large convection-driven precipitation, since the choice of the parametrization schemes and their tuning plays an essential role. The situation for the temperature field 
is similar but less uniformly so: while in Fig.~\ref{Figure_2} all the results cluster above the diagonal but roughly below WD~$\simeq 0.2$, the scatter in Fig.~\ref{Figure_3} is larger, with some results below the diagonal and some between $0.2 \lessapprox {\mathrm{WD}} \lessapprox 0.3$.

\begin{figure}
\noindent\includegraphics[width=\textwidth]{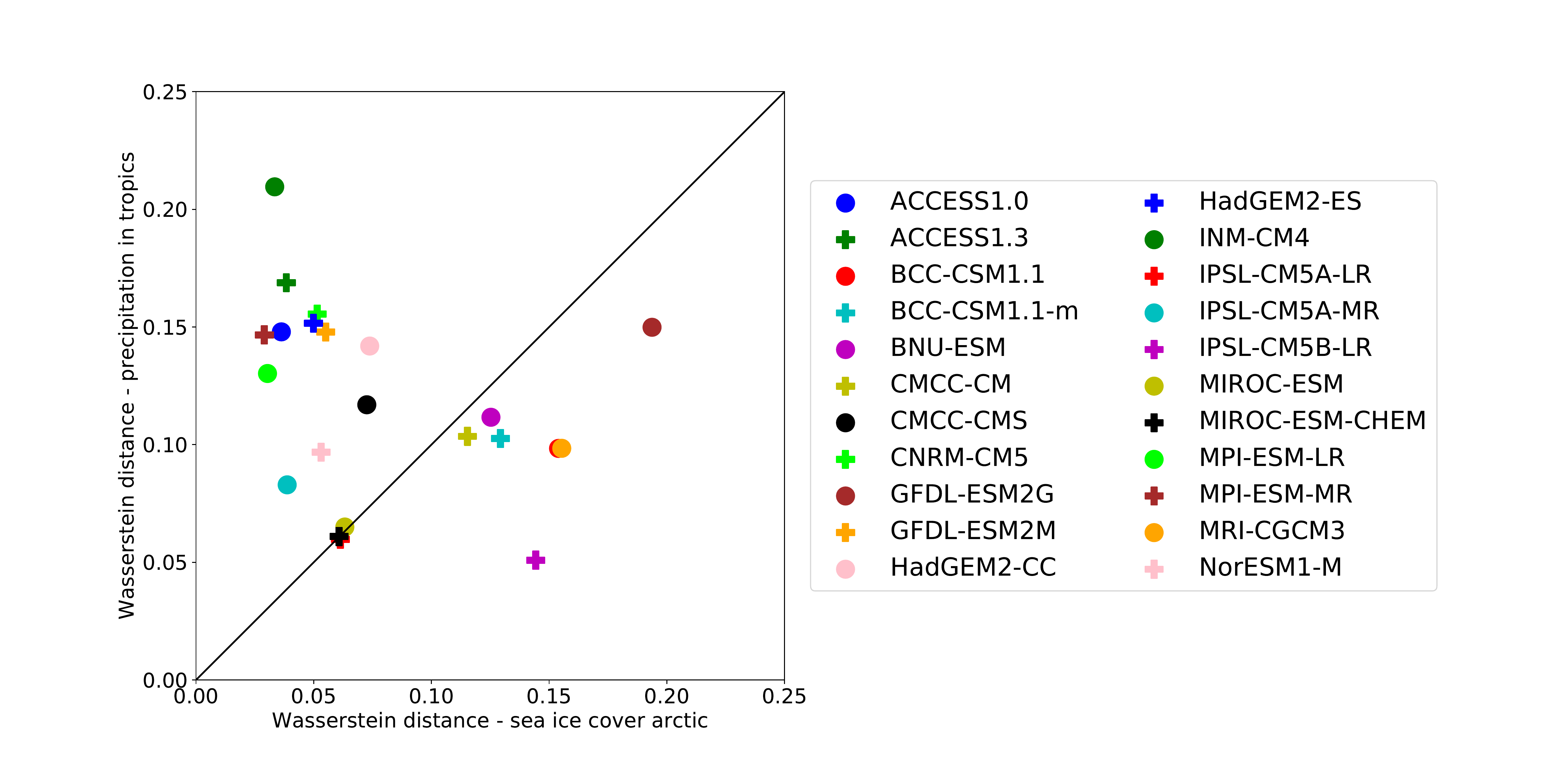}
\caption{One-dimensional WDs of average precipitation in the Tropics vs the average sea ice extent in the Arctic.}
\label{Figure_4}
\end{figure}

Figure~\ref{Figure_4} shows the scatter diagram of one-dimensional WDs for the precipitation in the Tropics vs the WDs of sea ice extent in the Arctic. Arctic sea ice cover is a very important indicator of the state of both hydrosphere and cryosphere, as well as of their mutual coupling; it is overestimated in CMIP5 models during the winter and spring seasons \cite{Randall2007, Flato2013}. 

Figure~\ref{Figure_4} demonstrates that the sea ice cover in the models is closer to the observations than the tropical precipitation in 12 CMIP5 models out of the 22 examined. Nevertheless, 7 models better describe tropical precipitation than sea ice extent in the Arctic, while 3 models have a similar --- and relatively low --- WD for both fields. This test indicates that a correct representation of the statistics of these two fields is  quite challenging for the CMIP5 models. 

\begin{figure}
 \noindent\includegraphics[width=\textwidth]{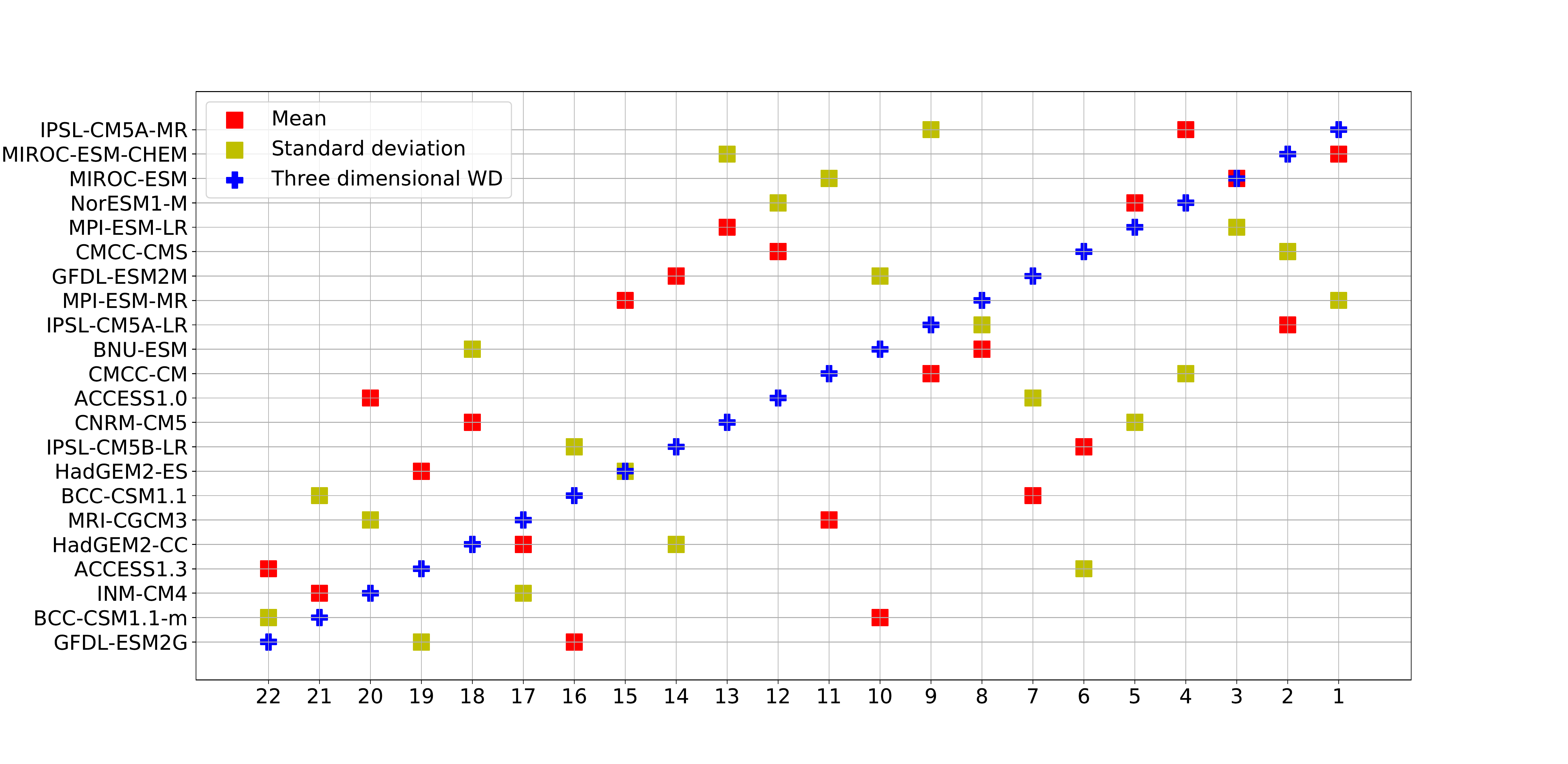}
\caption{Comparing 22 CMIP5 models (vertical axis) vs their positions in the ranking (horizontal axis):
     (a) three-dimensional WD -- heavy blue `$+$' sign; {\color{black}(b) mean -- red filled square; and 
     (c) standard deviation -- yellow filled square.} See text for explanations. See Tables S2-S4 in the Supplementary Information (SI) for detailed results.}
\label{Comparison}
\end{figure}

We compare next the performance of the CMIP5 models with respect to three different rankings. First, the three-dimensional WD is computed taking into account three physical quantities: globally averaged temperature and precipitation, along with sea ice extent in the Arctic. Note that, to ease the interpretation of Fig.~\ref{Comparison}, the models are listed on the vertical axis according to the rank provided by this methodology. 

The model ranking introduced herein is further compared with the rankings based on the first two moments of the distribution of reference. For each of the three physical quantities above, we compute the normalized mean, taking the absolute value of the difference between the mean of the distribution of the model field and that of the reference field, and dividing this difference by the standard deviation of the distribution of reference. The three means for the three fields are then averaged and the same procedure is repeated for the normalized standard deviation. 

We can see that the models' performance is quite different depending on the ranking being used. As an example, we focus on the BCC-CSM1.1 and BCC-CSM1.1-m models. The ranking based on the mean shows a rather good performance for both, with positions 7 and 10, respectively; nevertheless, they occupy positions 16 and 21 in the WD ranking. The latter low positions are due to their bad performances when it comes to standard deviation, where the two come last. 

The reverse instance is also clear by looking at those models that, while performing well in terms of variability, occupy lower rankings based on the WD due to their poor performance in the mean; see, for instance, the case of MPI-ESM-MR, with position 1 in the standard deviation, 8 in WD, and 15 in the mean. 
The WD score accounts for the information carried by the whole distribution --- i.e., by the mean, standard deviation and higher moments --- and clearly balances out the first and second moment thereof.

A more peculiar instance is provided by HadGEM2-CC and HadGEM2-ES, which rank in this order for both the mean (17th and 19th) and the standard deviation (14th and 15th), but in the reverse order in the WD ranking (18th and 15th). This apparent paradox could be due to the presence of nontrivial second-order correlations between the variables or from the effect of higher moments of the distributions.

Note that, for the 18-year time interval studied herein (1997--2014), the results obtained applying the WD approach in three-dimensional phase space are not very different from those given by averaging the three corresponding one-dimensional distances. This agreement is due to the unimodality of the distributions taken into account and things would be different in the case of multimodal distributions. 
In any case, the full application of the multi-dimensional WD leads to more robust results, as all correlations between the variables are taken into consideration.

\section{Conclusions} \label{sec:conclude}

We have proposed a new methodology to study the performance of climate models based on the computation of the Wasserstein distance (WD) between the multidimensional distributions of suitably chosen climatic fields of reference datasets and those of the models of interest. This method  takes into account all the moments of the distributions, {\color{black}unlike most evaluation methods for climate models used so far, which consider solely the first two moments of the distribution.} 
It is, therefore, more informative and takes into account also the distribution of extreme events. The methodology allows one to consider several variables at the same time, and it helps select such variables depending on the goal of the intercomparison. Thus, it can assist in disentangling the correlation between different climatic quantities. 

The proposed methodology has been proven to be effective in pointing to climate modeling problems related to the representation of quantities like precipitation or sea ice extent over limited areas, 
such as the Tropics and the Arctic, respectively; see again Figs.~\ref{Figure_2} and \ref{Figure_3}. Furthermore, this methodology can be 
applied to studying model performance for a given climatic variable over different spatial domains, as seen in Figs.~\ref{Figure_1}--\ref{Figure_4}, as well as relative model performance for different fields, as seen in Fig.~\ref{Figure_4}. This flexibility can help guide attempts at model improvements by providing robust diagnostics of the least well simulated field --- temperature, precipitation or sea ice extent --- or region, namely either hemisphere, the Tropics or the Arctic. 


Throughout the paper, we have shown the application of this approach to different physical fields, providing a ranking of CMIP5 models for specific sets of fields, as well as a way to highlight model weaknesses to help focus the honing of climate models. Getting more reliable models will lead to better simulations and, therefore, to more accurate climate predictions. 

\acknowledgments
GPCP and NCEP Reanalysis \nocite{Gibbs2002,KL.div.1951,Halmos.2017}2 data have been provided by NOAA/OAR/ESRL PSD 
at \url{https://www.esrl.noaa.gov/psd/}. The EUMETSAT 
OSI-SAF 
1979-2015 sea ice concentration 
(v2.0, 2017), {\color{black}has been provided by ICDC,  
University of Hamburg, at  \url{https://icdc.cen.uni-hamburg.de/en/seaiceconcentration-osisaf.html}. 
The authors thank the climate modeling groups for producing and making available their model output and acknowledge the World Climate Research Programme Working Group on Coupled Modelling. CMIP5 datasets are accessible at \url{https://esgf-data.dkrz.de}. {\color{black}The original data obtained in this paper are available at \url{https://figshare.com/articles/dataset/VissioLemboLucariniGhil2020_zip/12982406}. The authors thank 
E. Foufoula-Georgiou, T. T. Georgiou, and C. Kuehn for bringing to their attention relevant papers. 
VLu thanks N. Gigli and F. Santambrogio for inspiring exchanges. VLu and VLe have been supported by the DFG-CRC TRR181 (project no. 274762653).}
VLu and MG acknowledge the support received from the H2020 project TiPES (grant no. 820970). The present paper is TiPES contribution \#30. 
Work on this paper has also been supported by the EIT Climate-KIC (grant no. 190733)
}.

\section*{Appendix - Supporting Information}

\subsection*{Contents}
{\color{black}
\begin{enumerate}
\item Texts S1, S2, and S3;
\item Figure 1;
\item Tables S1 to S4.
\end{enumerate}}
%


\subsection*{Text S1. Wasserstein distance (WD): Background and history}

We present herein historical and mathematical information on WD, as well as additional information on the climate models analyzed. We wish to quantify the discrepancies between the output of a climate model and the observed reality by comparing their complete probability distributions and not just some representative quantity, like their variance. One way of doing so is to use the Kullback-Leibler (KL) divergence \cite{KL.div.1951}, which is rather widespread in applied statistics. To better explain the difference between the WD --- also called Monge-Kantorovich or Kantorovich-Rubinstein distance \cite{Kantorovich.1942} --- and the KL divergence, we first list below the axioms associated with the mathematical concept of a metric $d$. These axioms are inspired by and, of course, satisfied by the usual Euclidean distance. {\color{black}Note that, apart from WD, other metrics can be used for studying the distance between measures \cite{Gibbs2002}.}

Given points $x, y, z$ in a topological space $X$, $x, y, z \in X$, these axioms are 
\begin{subequations}\label{eq:metric}
\begin{align}
& d(x,y) = 0 \iff x = y, \label{eq:id} \\
& d(x,y) = d(y,x), \label{eq:symm} \\
& d(x,y) \le d(x,z) + d(z,y); \label{eq:triangle} 
\end{align}
\end{subequations}
they are referred to, respectively, as the axiom of identity or indiscernibles; the axiom of symmetry; and the axiom of subadditivity, better known as the triangle inequality. These axioms also imply the nonnegativity or separation condition $$d(x,y) \ge 0 \quad \mathrm{for~all} \quad x,y \in X.$$ 

A topological space $X$ equipped with such a metric becomes a metric space. Examples well-known in studying partial differential equations of fluid dynamics are so-called Hilbert spaces, which can be seen essentially as infinitely dimensional versions of Euclidean spaces \cite{Halmos.2017}.

Given probability distributions $P,Q,R$ on a metric space $X$, the KL divergence $D_{KL}(P\|Q)$ for $P$ given $Q$ satisfies neither the symmetry condition \eqref{eq:symm} nor the triangle inequality \eqref{eq:triangle}, i.e. 
\begin{subequations}\label{eq:KL}
\begin{align}
& D_{KL}(P\|Q) \neq D_{KL}(Q\|P)  \quad \mathrm{and, in~general,} \label{eq:asymm} \\
& D_{KL}(R\|P) \le D_{KL}(Q\|P) + D_{KL}(R\|Q) \quad \mathrm{does~not~hold}. \label{eq:nosub}
\end{align}
\end{subequations}

The WD \cite{Dobrushin.1970}, though, is a true metric and satisfies all three axioms of Eq.~\eqref{eq:metric}. It is based on the concept of optimal transport \cite{Villani2009} and it allows one to evaluate quantitatively the distance between two distributions: intuitively, the nearer the two distributions of points in phase space, the smaller the effort required to merge the two. 
WD is also called the ``earth mover's distance,'' since it was originally motivated by minimizing the effort of a platoon having dug a trench of prescribed shape and moving the earth dug up to another, existing trench of a different shape \cite{Monge1781}.

Using WD, it is possible to estimate the reliability of a model by choosing an appropriate combination of climatic or other physical variables, depending on the goal of the computation. Since an $N$-dimensional distribution contains much more information than its $N$ one-dimensional marginals, every point in our multidimensional distribution carries information about all the fields at the same time and not just 
about the product of the marginals.

\subsection*{Text S2. CMIP5 models}

The models that participated in CMIP5 are listed in Table~S1 below. The three rankings summarized in Fig.~5 of the Main Text are listed here in Tables~S2--S4.

\renewcommand{\thetable}{S\arabic{table}}

 \begin{sidewaystable}
 \centering
 \resizebox{\textwidth}{!}{
  \begin{tabular}{|c|c|c|c|}
  \hline
  Acronym & Model & Center & Country \\  [0.5ex] 
  \hline
  ACCESS1.0 & Australian Community Climate and Earth-System Simulator, version 1.0 & Commonwealth Scientific and Industrial Research Organisation – Bureau of Meteorology  & Australia \\
  ACCESS1.3 & Australian Community Climate and Earth-System Simulator, version 1.3 & Commonwealth Scientific and Industrial Research Organisation – Bureau of Meteorology (CSIRO-BOM) & Australia \\
  BCC-CSM1.1 & Beijing Climate Center, Climate System Model, version 1.1 & Beijing Climate Center (BCC), Chinese Meteorological Administration (CMA) & China \\
  BCC-CSM1.1-m & Beijing Climate Center, Climate System Model, version 1.1, Moderate resolution & Beijing Climate Center (BCC), Chinese Meteorological Administration (CMA) & China \\
  BNU-ESM & Beijing Normal University Earth System Model & Beijing Normal University (BNU) & China \\
  CanESM2* & Second Generation Canadian Earth System Model & Canadian Centre for Climate Modelling and Analysis (CCCma) & Canada \\
  CCSM4* & Community Climate System Model, version 4 & National Center for Atmospheric Research (NCAR) & United States of America \\
  CESM1-BGC* & Community Earth System Model, version 1, BioGeoChemistry & National Science Foundation (NSF) – U.S. Department of Energy (DOE) – National Center for Atmospheric Research (NCAR) & United States of America \\
  CESM1-CAM5* & Community Earth System Model, version 1 - Community Atmosphere Model, version 5 & National Science Foundation (NSF) – U.S. Department of Energy (DOE) – National Center for Atmospheric Research (NCAR) & United States of America \\
  CMCC-CM & Centro Euro-Mediterraneo per I Cambiamenti Climatici Climate Model & Centro Euro-Mediterraneo per I Cambiamenti Climatici (CMCC) & Italy \\
  CMCC-CMS & Centro Euro-Mediterraneo per I Cambiamenti Climatici Climate Model, Stratosphere version & Centro Euro-Mediterraneo per I Cambiamenti Climatici (CMCC) & Italy \\
  CNRM-CM5 & Centre National de Recherches Météorologiques Coupled Global Climate Model, version 5 & Centre National de Recherches Météorologiques (CNRM) – Centre Européen de Recherche et de Formation Avancée en Calcul Scientifique (CERFACS) & France \\
  CSIRO-Mk3.6.0* & Commonwealth Scientific and Industrial Research Organisation Mark, version 3.6.0 & Commonwealth Scientific and Industrial Research Organisation (CSIRO) – Queensland Climate Change Centre of Excellence (QCCCE) & Australia \\
  EC-EARTH* & European Community Earth-System Model & EC-EARTH Consortium & Europe \\
  GFDL-CM3* & Geophysical Fluid Dynamics Laboratory Climate Model, version 3 & National Oceanic and Atmospheric Administration (NOAA) - Geophysical Fluid Dynamics Laboratory (GFDL) & United States of America \\
  GFDL-ESM2G & Geophysical Fluid Dynamics Laboratory Earth System Model, Generalized Ocean Layer Dynamics (GOLD) component & National Oceanic and Atmospheric Administration (NOAA) - Geophysical Fluid Dynamics Laboratory (GFDL) & United States of America \\
  GFDL-ESM2M & Geophysical Fluid Dynamics Laboratory Earth System Model, Modular Ocean Model 4 (MOM4) component & National Oceanic and Atmospheric Administration (NOAA) - Geophysical Fluid Dynamics Laboratory (GFDL) & United States of America \\
  HadGEM2-CC & Hadley Centre Global Environment Model, version 2, Carbon Cycle & Met Office Hadley Centre & United Kingdom \\
  HadGEM2-ES & Hadley Centre Global Environment Model, version 2, Earth System & Met Office Hadley Centre & United Kingdom \\
  INM-CM4 & Institute of Numerical Mathematics Coupled Model, version 4.0 & Institute of Numerical Mathematics (INM) & Russia \\
  IPSL-CM5A-LR & Institut Pierre-Simon Laplace Coupled Model, version 5A, Low Resolution & Institut Pierre-Simon Laplace (IPSL) & France \\
  IPSL-CM5A-MR & Institut Pierre-Simon Laplace Coupled Model, version 5A, Medium Resolution & Institut Pierre-Simon Laplace (IPSL) & France \\
  IPSL-CM5B-LR & Institut Pierre-Simon Laplace Coupled Model, version 5B, Low Resolution & Institut Pierre-Simon Laplace (IPSL) & France \\
  MIROC5* & Model for Interdisciplinary Research on Climate, version 5 & Atmosphere and Ocean Research Institute (AORI)–National Institute for Environmental Studies (NIES)–Japan Agency for Marine-Earth Science and Technology (JAMSTEC) & Japan \\
  MIROC-ESM & Model for Interdisciplinary Research on Climate, Earth System Model & Atmosphere and Ocean Research Institute (AORI)–National Institute for Environmental Studies (NIES)–Japan Agency for Marine-Earth Science and Technology (JAMSTEC) & Japan \\
  MIROC-ESM-CHEM & Model for Interdisciplinary Research on Climate, Earth System Model, atmospheric chemistry coupled version & Atmosphere and Ocean Research Institute (AORI)–National Institute for Environmental Studies (NIES)–Japan Agency for Marine-Earth Science and Technology (JAMSTEC) & Japan \\
  MPI-ESM-LR & Max Planck Institute Earth System Model, Low Resolution & Max Planck Institute for Meteorology (MPI-M) & Germany \\
  MPI-ESM-MR & Max Planck Institute Earth System Model, Medium Resolution & Max Planck Institute for Meteorology (MPI-M) & Germany \\
  MRI-CGCM3 & Meteorological Research Institute Coupled Atmosphere–Ocean General Circulation Model, version 3 & Meteorological Research Institute (MRI) & Japan \\
  NorESM1-M & Norwegian Earth System Model, version 1, Medium Resolution & Norwegian Climate Centre (NCC) & Norway \\ [1ex] 
  \hline
 \end{tabular}}
 \caption{CMIP5 models used in the paper. The asterisk points out the models not used for sea ice entension tests.}
 \label{table1}
 \end{sidewaystable}
 
 \begin{table}
 \centering
  \begin{tabular}{|c|c|}
  \hline
  3D WD & Model \\ [0.5ex] 
  \hline
  0.097 & IPSL-CM5A-MR \\ 
  0.101 & MIROC-ESM-CHEM \\
  0.107 & MIROC-ESM \\
  0.125 & NorESM1-M \\
  0.136 & MPI-ESM-LR \\  
  0.143 & CMCC-CMS \\
  0.157 & GFDL-ESM2M \\
  0.158 & MPI-ESM-MR \\
  0.162 & IPSL-CM5A-LR \\
  0.165 & BNU-ESM \\
  0.169 & CMCC-CM \\
  0.188 & ACCESS1.0 \\
  0.188 & CNRM-CM5 \\
  0.191 & IPSL-CM5B-LR \\
  0.192 & HadGEM2-ES \\
  0.193 & BCC-CSM1.1 \\
  0.200 & MRI-CGCM3 \\
  0.207 & HadGEM2-CC \\
  0.223 & ACCESS1.3 \\
  0.229 & INM-CM4 \\
  0.235 & BCC-CSM1.1-m \\
  0.246 & GFDL-ESM2G \\ [1ex]
  \hline
 \end{tabular}
 \caption{Ranking of CMIP5 models obtained with the three-dimensional WD.}
 \label{table2}
 \end{table}

 \begin{table}
 \centering
 \begin{tabular}{|c|c|}
  \hline
Average of means & Model \\ [0.5ex] 
  \hline
  0.881 & MIROC-ESM-CHEM \\
  0.978 & IPSL-CM5A-LR \\
  0.993 & MIROC-ESM \\
  1.030 & IPSL-CM5A-MR \\
  1.128 & NorESM1-M \\
  1.369 & IPSL-CM5B-LR \\
  1.412 & BCC-CSM1.1 \\
  1.557 & BNU-ESM \\
  1.748 & CMCC-CM \\
  1.749 & BCC-CSM1.1-m \\
  1.785 & MRI-CGCM3 \\
  1.785 & CMCC-CMS \\
  1.893 & MPI-ESM-LR \\
  2.120 & GFDL-ESM2M \\
  2.224 & MPI-ESM-MR \\
  2.335 & GFDL-ESM2G \\
  2.508 & HadGEM2-CC \\
  2.578 & CNRM-CM5 \\
  2.657 & HadGEM2-ES \\
  2.694 & ACCESS1.0 \\
  3.163 & INM-CM4 \\
  3.239 & ACCESS1.3 \\ [1ex]
  \hline
 \end{tabular}
 \caption{Ranking obtained by averaging the three separate mean distances.}
 \label{table3}
 \end{table}

 \begin{table}
 \centering
  \begin{tabular}{|c|c|} 
  \hline
  Average of the \\standard deviations & Model \\ [0.5ex] 
  \hline
  0.160 & MPI-ESM-MR \\
  0.186 & CMCC-CMS \\
  0.189 & MPI-ESM-LR \\
  0.225 & CMCC-CM \\
  0.298 & CNRM-CM5 \\
  0.326 & ACCESS1.3 \\
  0.360 & ACCESS1.0 \\
  0.362 & IPSL-CM5A-LR \\
  0.366 & IPSL-CM5A-MR \\
  0.369 & GFDL-ESM2M \\
  0.390 & MIROC-ESM \\
  0.391 & NorESM1-M \\
  0.406 & MIROC-ESM-CHEM \\
  0.434 & HadGEM2-CC \\
  0.443 & HadGEM2-ES \\
  0.452 & IPSL-CM5B-LR \\
  0.455 & INM-CM4 \\
  0.532 & BNU-ESM \\
  0.573 & GFDL-ESM2G \\
  0.651 & MRI-CGCM3 \\
  0.758 & BCC-CSM1.1 \\
  0.762 & BCC-CSM1.1-m \\ [1ex]
  \hline
 \end{tabular}
 \caption{Ranking obtained by averaging the three standard deviations.}
 \label{table4}
 \end{table}

\begin{figure}
\noindent\includegraphics[angle=270,width=\textwidth]{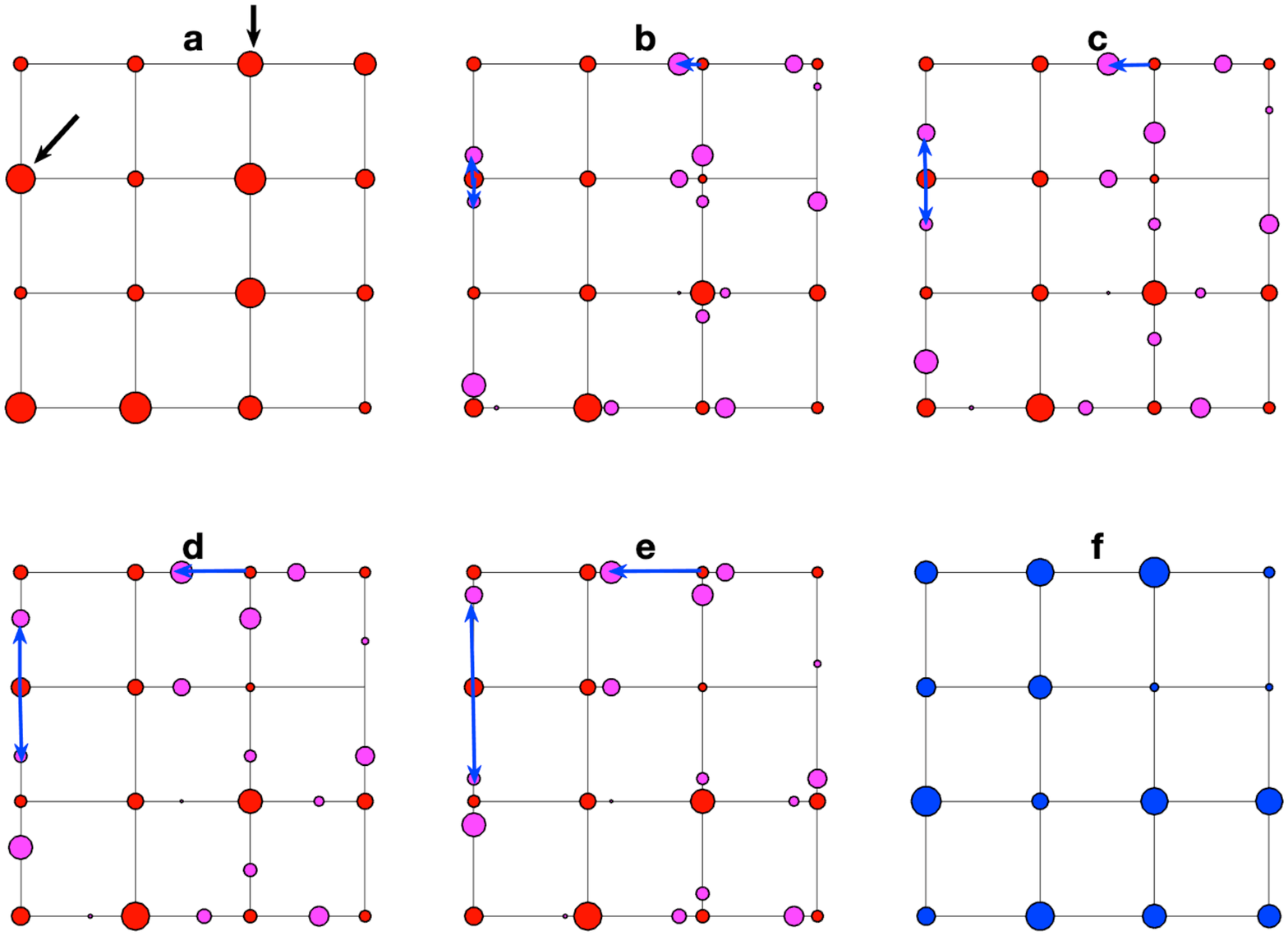}
\caption{Six consecutive steps of mass transport from the initial measure in panel (a) to the final one in panel (f). {\color{black} The red  circles represent the initial measure in panel (a) and the mass that is not moved by the optimal transport protocol in panels (b)--(e). The black  circles indicate the mass being transported by the protocol towards the target distribution; see blue circles in panel (f).  The size of the circles is proportional to the corresponding mass. The blue arrows in panels (b)--(e) indicate the mass movement away from the two grid points marked by the black arrows in panel (a).}} 
\label{transport}
\end{figure}

{\color{black}\subsection*{Text S3. Optimal transport}

To facilitate the understanding of the methodology presented herein, we show in Fig.~\ref{transport} the consecutive steps used in merging a distribution, shown in step (a), into another distribution, shown in step (f). The square grid of $N \times N$ in the Main Text uses $N = 20$; this size is obtained by a trade-off between the conflicting requirements of accuracy --- the larger $N$ the better --- and computability --- the smaller $N$ the better. For clarity, we use here $N = 4$. 
 For illustrative purposes, we indicate explicitly how the optimal transport protocol moves mass --- according to the blue arrows in panels (b)-(e) --- away from the two grid points marked by the black arrows in panel (a). In general, the mass contained in the different nodes is shuffled around so that the morphing from the initial to the final measure is achieved  with the least possible effort.}

\newpage


%
%


\begin{thebibliography}{}

\bibitem [\protect \citeauthoryear {%
Adler%
\ \protect \BOthers {.}}{%
Adler%
\ \protect \BOthers {.}}{%
{\protect \APACyear {2003}}%
}]{%
Adler2003}
\APACinsertmetastar {%
Adler2003}%
\begin{APACrefauthors}%
Adler, R\BPBI F.%
, Huffman, G\BPBI J.%
, Chang, A.%
, Ferraro, R.%
, Xie, P.%
, Janowiak, J.%
\BDBL {}Arkin, P.%
\end{APACrefauthors}%
\unskip\
\newblock
\APACrefYearMonthDay{2003}{}{}.
\newblock
{\BBOQ}\APACrefatitle {The {Version 2 Global Precipitation Climatology Project
  (GPCP)} monthly precipitation analysis (1979--present)} {The {Version 2
  Global Precipitation Climatology Project (GPCP)} monthly precipitation
  analysis (1979--present)}.{\BBCQ}
\newblock
\APACjournalVolNumPages{Journal of Hydrometeorology}{4}{}{1147--1167}.
\PrintBackRefs{\CurrentBib}

\bibitem [\protect \citeauthoryear {%
Dee%
\ \protect \BOthers {.}}{%
Dee%
\ \protect \BOthers {.}}{%
{\protect \APACyear {2011}}%
}]{%
Dee2011}
\APACinsertmetastar {%
Dee2011}%
\begin{APACrefauthors}%
Dee, D\BPBI P.%
, Uppala, S\BPBI M.%
, Simmons, A\BPBI J.%
, Berrisford, P.%
, Poli, P.%
, Kobayashi, S.%
\BDBL {}Vitart, F.%
\end{APACrefauthors}%
\unskip\
\newblock
\APACrefYearMonthDay{2011}{}{}.
\newblock
{\BBOQ}\APACrefatitle {The {ERA-Interim reanalysis: configuration and
  performance of the data assimilation system}} {The {ERA-Interim reanalysis:
  configuration and performance of the data assimilation system}}.{\BBCQ}
\newblock
\APACjournalVolNumPages{Quarterly Journal of the Royal Meteorological
  Society}{137}{656}{553-597}.
\PrintBackRefs{\CurrentBib}

\bibitem [\protect \citeauthoryear {%
Dobrushin%
}{%
Dobrushin%
}{%
{\protect \APACyear {1970}}%
}]{%
Dobrushin.1970}
\APACinsertmetastar {%
Dobrushin.1970}%
\begin{APACrefauthors}%
Dobrushin, R\BPBI L.%
\end{APACrefauthors}%
\unskip\
\newblock
\APACrefYearMonthDay{1970}{}{}.
\newblock
{\BBOQ}\APACrefatitle {Prescribing a system of random variables by conditional
  distributions} {Prescribing a system of random variables by conditional
  distributions}.{\BBCQ}
\newblock
\APACjournalVolNumPages{Theory of Probability \& Its
  Applications}{15}{3}{458--486}.
\PrintBackRefs{\CurrentBib}

\bibitem [\protect \citeauthoryear {%
{EUMETSAT Ocean and Sea Ice Satellite Application Facility}%
}{%
{EUMETSAT Ocean and Sea Ice Satellite Application Facility}%
}{%
{\protect \APACyear {2017}}%
}]{%
EUMETSAT2017}
\APACinsertmetastar {%
EUMETSAT2017}%
\begin{APACrefauthors}%
{EUMETSAT Ocean and Sea Ice Satellite Application Facility}.%
\end{APACrefauthors}%
\unskip\
\newblock
\APACrefYearMonthDay{2017}{}{}.
\newblock
\APACrefbtitle {{Global sea ice concentration climate data record 1979-2015
  (v2.0, 2017). Norwegian and Danish Meteorological Institutes.}} {{Global sea
  ice concentration climate data record 1979-2015 (v2.0, 2017). Norwegian and
  Danish Meteorological Institutes.}}
\newblock
\APAChowpublished {\url{http://osisaf.met.no.}}
\newblock
\begin{APACrefDOI} \doi{10.15770/EUM_SAF_OSI_0008} \end{APACrefDOI}
\PrintBackRefs{\CurrentBib}

\bibitem [\protect \citeauthoryear {%
Eyring%
\ \protect \BOthers {.}}{%
Eyring%
\ \protect \BOthers {.}}{%
{\protect \APACyear {2020}}%
}]{%
Eyring2020}
\APACinsertmetastar {%
Eyring2020}%
\begin{APACrefauthors}%
Eyring, V.%
, Bock, L.%
, Lauer, A.%
, Righi, M.%
, Schlund, M.%
, Andela, B.%
\BDBL {}Zimmermann, K.%
\end{APACrefauthors}%
\unskip\
\newblock
\APACrefYearMonthDay{2020}{}{}.
\newblock
{\BBOQ}\APACrefatitle {Earth System Model Evaluation Tool (ESMValTool) v2.0 --
  an extended set of large-scale diagnostics for quasi-operational and
  comprehensive evaluation of Earth system models in CMIP} {Earth system model
  evaluation tool (esmvaltool) v2.0 -- an extended set of large-scale
  diagnostics for quasi-operational and comprehensive evaluation of earth
  system models in cmip}.{\BBCQ}
\newblock
\APACjournalVolNumPages{Geoscientific Model Development}{13}{7}{3383--3438}.
\newblock
\begin{APACrefURL} \url{https://gmd.copernicus.org/articles/13/3383/2020/}
  \end{APACrefURL}
\newblock
\begin{APACrefDOI} \doi{10.5194/gmd-13-3383-2020} \end{APACrefDOI}
\PrintBackRefs{\CurrentBib}

\bibitem [\protect \citeauthoryear {%
Eyring%
, Bony%
\BCBL {}\ \protect \BOthers {.}}{%
Eyring%
, Bony%
\BCBL {}\ \protect \BOthers {.}}{%
{\protect \APACyear {2016}}%
}]{%
Eyring2016a}
\APACinsertmetastar {%
Eyring2016a}%
\begin{APACrefauthors}%
Eyring, V.%
, Bony, S.%
, Meehl, G\BPBI A.%
, Senior, C\BPBI A.%
, Stevens, B.%
, Stouffer, R\BPBI J.%
\BCBL {}\ \BBA {} Taylor, K\BPBI E.%
\end{APACrefauthors}%
\unskip\
\newblock
\APACrefYearMonthDay{2016}{}{}.
\newblock
{\BBOQ}\APACrefatitle {{Overview of the Coupled Model Intercomparison Project
  Phase 6 (CMIP6) experimental design and organization}} {{Overview of the
  Coupled Model Intercomparison Project Phase 6 (CMIP6) experimental design and
  organization}}.{\BBCQ}
\newblock
\APACjournalVolNumPages{Geoscientific Model Development}{9}{}{10539--10583}.
\PrintBackRefs{\CurrentBib}

\bibitem [\protect \citeauthoryear {%
Eyring%
, Righi%
\BCBL {}\ \protect \BOthers {.}}{%
Eyring%
, Righi%
\BCBL {}\ \protect \BOthers {.}}{%
{\protect \APACyear {2016}}%
}]{%
Eyring2016b}
\APACinsertmetastar {%
Eyring2016b}%
\begin{APACrefauthors}%
Eyring, V.%
, Righi, M.%
, Lauer, A.%
, Evaldsson, M.%
, Wenzel, S.%
, Jones, C.%
\BDBL {}Williams, K\BPBI D.%
\end{APACrefauthors}%
\unskip\
\newblock
\APACrefYearMonthDay{2016}{}{}.
\newblock
{\BBOQ}\APACrefatitle {{ESMValTool (v1.0) – a community diagnostic and
  performance metrics tool for routine evaluation of Earth system models in
  CMIP}} {{ESMValTool (v1.0) – a community diagnostic and performance metrics
  tool for routine evaluation of Earth system models in CMIP}}.{\BBCQ}
\newblock
\APACjournalVolNumPages{Geoscientific Model Development}{9}{}{1747--1802}.
\PrintBackRefs{\CurrentBib}

\bibitem [\protect \citeauthoryear {%
Flato%
\ \protect \BOthers {.}}{%
Flato%
\ \protect \BOthers {.}}{%
{\protect \APACyear {2013}}%
}]{%
Flato2013}
\APACinsertmetastar {%
Flato2013}%
\begin{APACrefauthors}%
Flato, G.%
, Marotzke, J.%
, Abiodun, B.%
, Braconnot, P.%
, Chou, S\BPBI C.%
, Collins, W.%
\BDBL {}Rummukainen, M.%
\end{APACrefauthors}%
\unskip\
\newblock
\APACrefYearMonthDay{2013}{}{}.
\newblock
{\BBOQ}\APACrefatitle {Evaluation of climate models} {Evaluation of climate
  models}.{\BBCQ}
\newblock
\BIn{} T.~Stocker\ \BOthers {.}\ (\BEDS), \APACrefbtitle {{Climate Change 2013:
  The Physical Science Basis. Contribution of Working Group I to the Fifth
  Assessment Report of the Intergovernmental Panel on Climate Change}.}
  {{Climate Change 2013: The Physical Science Basis. Contribution of Working
  Group I to the Fifth Assessment Report of the Intergovernmental Panel on
  Climate Change}.}
\newblock
\APACaddressPublisher{Cambridge, UK and NY, USA}{Cambridge University Press}.
\PrintBackRefs{\CurrentBib}

\bibitem [\protect \citeauthoryear {%
Ghil%
}{%
Ghil%
}{%
{\protect \APACyear {2015}}%
}]{%
Ghil2015}
\APACinsertmetastar {%
Ghil2015}%
\begin{APACrefauthors}%
Ghil, M.%
\end{APACrefauthors}%
\unskip\
\newblock
\APACrefYearMonthDay{2015}{}{}.
\newblock
{\BBOQ}\APACrefatitle {A mathematical theory of climate sensitivity or, {How}
  to deal with both anthropogenic forcing and natural variability?} {A
  mathematical theory of climate sensitivity or, {How} to deal with both
  anthropogenic forcing and natural variability?}{\BBCQ}
\newblock
\BIn{} C\BHBI P.~Chang, M.~Ghil, M.~Latif\BCBL {}\ \BBA {} J.~Wallace\ (\BEDS),
  \APACrefbtitle {{Climate Change: Multidecadal and Beyond}} {{Climate Change:
  Multidecadal and Beyond}}\ (\BVOL~6, \BPG~31-52).
\newblock
\APACaddressPublisher{Singapore}{World Scientific Publishing Co.}
\PrintBackRefs{\CurrentBib}

\bibitem [\protect \citeauthoryear {%
Ghil%
\ \BBA {} Lucarini%
}{%
Ghil%
\ \BBA {} Lucarini%
}{%
{\protect \APACyear {2020}}%
}]{%
Ghil.Luc.2020}
\APACinsertmetastar {%
Ghil.Luc.2020}%
\begin{APACrefauthors}%
Ghil, M.%
\BCBT {}\ \BBA {} Lucarini, V.%
\end{APACrefauthors}%
\unskip\
\newblock
\APACrefYearMonthDay{2020}{Jul}{}.
\newblock
{\BBOQ}\APACrefatitle {The physics of climate variability and climate change}
  {The physics of climate variability and climate change}.{\BBCQ}
\newblock
\APACjournalVolNumPages{Rev. Mod. Phys.}{92}{}{035002}.
\newblock
\begin{APACrefURL} \url{https://link.aps.org/doi/10.1103/RevModPhys.92.035002}
  \end{APACrefURL}
\newblock
\begin{APACrefDOI} \doi{10.1103/RevModPhys.92.035002} \end{APACrefDOI}
\PrintBackRefs{\CurrentBib}

\bibitem [\protect \citeauthoryear {%
Gibbs%
\ \BBA {} Su%
}{%
Gibbs%
\ \BBA {} Su%
}{%
{\protect \APACyear {2002}}%
}]{%
Gibbs2002}
\APACinsertmetastar {%
Gibbs2002}%
\begin{APACrefauthors}%
Gibbs, A\BPBI L.%
\BCBT {}\ \BBA {} Su, F\BPBI E.%
\end{APACrefauthors}%
\unskip\
\newblock
\APACrefYearMonthDay{2002}{}{}.
\newblock
{\BBOQ}\APACrefatitle {On choosing and bounding probability metrics} {On
  choosing and bounding probability metrics}.{\BBCQ}
\newblock
\APACjournalVolNumPages{{International Statistical Review}}{70}{3}{419--435}.
\PrintBackRefs{\CurrentBib}

\bibitem [\protect \citeauthoryear {%
Gleckler%
, Taylor%
\BCBL {}\ \BBA {} Doutriaux%
}{%
Gleckler%
\ \protect \BOthers {.}}{%
{\protect \APACyear {2008}}%
}]{%
Gleckler2008}
\APACinsertmetastar {%
Gleckler2008}%
\begin{APACrefauthors}%
Gleckler, P\BPBI J.%
, Taylor, K\BPBI E.%
\BCBL {}\ \BBA {} Doutriaux, C.%
\end{APACrefauthors}%
\unskip\
\newblock
\APACrefYearMonthDay{2008}{}{}.
\newblock
{\BBOQ}\APACrefatitle {{Performance metrics for climate models}} {{Performance
  metrics for climate models}}.{\BBCQ}
\newblock
\APACjournalVolNumPages{Journal of Geophysical Research}{113}{}{D06104}.
\PrintBackRefs{\CurrentBib}

\bibitem [\protect \citeauthoryear {%
Halmos%
}{%
Halmos%
}{%
{\protect \APACyear {2017}}%
}]{%
Halmos.2017}
\APACinsertmetastar {%
Halmos.2017}%
\begin{APACrefauthors}%
Halmos, P\BPBI R.%
\end{APACrefauthors}%
\unskip\
\newblock
\APACrefYear{2017}.
\newblock
\APACrefbtitle {{Introduction to Hilbert Space and the Theory of Spectral
  Multiplicity}} {{Introduction to Hilbert Space and the Theory of Spectral
  Multiplicity}}.
\newblock
\APACaddressPublisher{}{Courier Dover Publications}.
\PrintBackRefs{\CurrentBib}

\bibitem [\protect \citeauthoryear {%
Held%
}{%
Held%
}{%
{\protect \APACyear {2005}}%
}]{%
Held2005}
\APACinsertmetastar {%
Held2005}%
\begin{APACrefauthors}%
Held, I\BPBI M.%
\end{APACrefauthors}%
\unskip\
\newblock
\APACrefYearMonthDay{2005}{}{}.
\newblock
{\BBOQ}\APACrefatitle {The gap between simulation and understanding in climate
  modeling} {The gap between simulation and understanding in climate
  modeling}.{\BBCQ}
\newblock
\APACjournalVolNumPages{Bulletin of the American Meteorological
  Society}{86}{}{1609–1614}.
\PrintBackRefs{\CurrentBib}

\bibitem [\protect \citeauthoryear {%
Kanamitsu%
\ \protect \BOthers {.}}{%
Kanamitsu%
\ \protect \BOthers {.}}{%
{\protect \APACyear {2002}}%
}]{%
Kanamitsu2002}
\APACinsertmetastar {%
Kanamitsu2002}%
\begin{APACrefauthors}%
Kanamitsu, M.%
, Ebisuzaki, W.%
, Woollen, J.%
, Yang, S\BHBI K.%
, Hnilo, J\BPBI J.%
, Fiorino, M.%
\BCBL {}\ \BBA {} Potter, G\BPBI L.%
\end{APACrefauthors}%
\unskip\
\newblock
\APACrefYearMonthDay{2002}{}{}.
\newblock
{\BBOQ}\APACrefatitle {{NCEP-DOE AMIP-II Reanalysis (R-2)}} {{NCEP-DOE AMIP-II
  Reanalysis (R-2)}}.{\BBCQ}
\newblock
\APACjournalVolNumPages{Bulletin of the American Meteorological Society}{Nov
  2002}{}{1631-1643}.
\PrintBackRefs{\CurrentBib}

\bibitem [\protect \citeauthoryear {%
Kantorovich%
}{%
Kantorovich%
}{%
{\protect \APACyear {2006}}%
}]{%
Kantorovich.1942}
\APACinsertmetastar {%
Kantorovich.1942}%
\begin{APACrefauthors}%
Kantorovich, L\BPBI V.%
\end{APACrefauthors}%
\unskip\
\newblock
\APACrefYearMonthDay{2006}{}{}.
\newblock
{\BBOQ}\APACrefatitle {On the translocation of masses} {On the translocation of
  masses}.{\BBCQ}
\newblock
\APACjournalVolNumPages{Journal of Mathematical Sciences}{133}{4}{1381--1382}.
\newblock
\APACrefnote{{originally published in Doklady Akademii Nauk SSSR, 37 (7–8),
  199--201 (1942).}}
\PrintBackRefs{\CurrentBib}

\bibitem [\protect \citeauthoryear {%
Knutti%
\ \protect \BOthers {.}}{%
Knutti%
\ \protect \BOthers {.}}{%
{\protect \APACyear {2017}}%
}]{%
Knutti2017}
\APACinsertmetastar {%
Knutti2017}%
\begin{APACrefauthors}%
Knutti, R.%
, Sedl{\'{a}}{\v{c}}ek, J.%
, Sanderson, B\BPBI M.%
, Lorenz, R.%
, Fischer, E\BPBI M.%
\BCBL {}\ \BBA {} Eyring, V.%
\end{APACrefauthors}%
\unskip\
\newblock
\APACrefYearMonthDay{2017}{}{}.
\newblock
{\BBOQ}\APACrefatitle {{A climate model projection weighting scheme accounting
  for performance and interdependence}} {{A climate model projection weighting
  scheme accounting for performance and interdependence}}.{\BBCQ}
\newblock
\APACjournalVolNumPages{Geophysical Research Letters}{44}{}{1909--1918}.
\PrintBackRefs{\CurrentBib}

\bibitem [\protect \citeauthoryear {%
Kullback%
\ \BBA {} Leibler%
}{%
Kullback%
\ \BBA {} Leibler%
}{%
{\protect \APACyear {1951}}%
}]{%
KL.div.1951}
\APACinsertmetastar {%
KL.div.1951}%
\begin{APACrefauthors}%
Kullback, S.%
\BCBT {}\ \BBA {} Leibler, R\BPBI A.%
\end{APACrefauthors}%
\unskip\
\newblock
\APACrefYearMonthDay{1951}{}{}.
\newblock
{\BBOQ}\APACrefatitle {On information and sufficiency} {On information and
  sufficiency}.{\BBCQ}
\newblock
\APACjournalVolNumPages{The Annals of Mathematical Statistics}{22}{1}{79--86}.
\PrintBackRefs{\CurrentBib}

\bibitem [\protect \citeauthoryear {%
Lembo%
, Lunkeit%
\BCBL {}\ \BBA {} Lucarini%
}{%
Lembo%
\ \protect \BOthers {.}}{%
{\protect \APACyear {2019}}%
}]{%
Lembo2019}
\APACinsertmetastar {%
Lembo2019}%
\begin{APACrefauthors}%
Lembo, V.%
, Lunkeit, F.%
\BCBL {}\ \BBA {} Lucarini, V.%
\end{APACrefauthors}%
\unskip\
\newblock
\APACrefYearMonthDay{2019}{}{}.
\newblock
{\BBOQ}\APACrefatitle {TheDiaTo (v1.0) -- a new diagnostic tool for water,
  energy and entropy budgets in climate models} {Thediato (v1.0) -- a new
  diagnostic tool for water, energy and entropy budgets in climate
  models}.{\BBCQ}
\newblock
\APACjournalVolNumPages{Geoscientific Model Development}{12}{8}{3805--3834}.
\newblock
\begin{APACrefURL} \url{https://gmd.copernicus.org/articles/12/3805/2019/}
  \end{APACrefURL}
\newblock
\begin{APACrefDOI} \doi{10.5194/gmd-12-3805-2019} \end{APACrefDOI}
\PrintBackRefs{\CurrentBib}

\bibitem [\protect \citeauthoryear {%
Lorenz%
}{%
Lorenz%
}{%
{\protect \APACyear {1984}}%
}]{%
Lorenz.1984}
\APACinsertmetastar {%
Lorenz.1984}%
\begin{APACrefauthors}%
Lorenz, E\BPBI N.%
\end{APACrefauthors}%
\unskip\
\newblock
\APACrefYearMonthDay{1984}{}{}.
\newblock
{\BBOQ}\APACrefatitle {Irregularity: {A fundamental property of the
  atmosphere}} {Irregularity: {A fundamental property of the
  atmosphere}}.{\BBCQ}
\newblock
\APACjournalVolNumPages{Tellus A}{36}{2}{98--110}.
\PrintBackRefs{\CurrentBib}

\bibitem [\protect \citeauthoryear {%
Lucarini%
}{%
Lucarini%
}{%
{\protect \APACyear {2013}}%
}]{%
Lucarini2013}
\APACinsertmetastar {%
Lucarini2013}%
\begin{APACrefauthors}%
Lucarini, V.%
\end{APACrefauthors}%
\unskip\
\newblock
\APACrefYearMonthDay{2013}{}{}.
\newblock
{\BBOQ}\APACrefatitle {{Modeling complexity: the case of climate science}}
  {{Modeling complexity: the case of climate science}}.{\BBCQ}
\newblock
\BIn{} U.~Gähde, S.~Hartmann\BCBL {}\ \BBA {} J.~Wolf\ (\BEDS), \APACrefbtitle
  {Models, Simulations, and the Reduction of Complexity} {Models, simulations,
  and the reduction of complexity}\ (\BPG~229-254).
\newblock
\APACaddressPublisher{}{De Gruyter}.
\PrintBackRefs{\CurrentBib}

\bibitem [\protect \citeauthoryear {%
Lucarini%
, Calmanti%
, Dell'Aquila%
, Ruti%
\BCBL {}\ \BBA {} Speranza%
}{%
Lucarini%
\ \protect \BOthers {.}}{%
{\protect \APACyear {2007}}%
}]{%
Lucarini2007}
\APACinsertmetastar {%
Lucarini2007}%
\begin{APACrefauthors}%
Lucarini, V.%
, Calmanti, S.%
, Dell'Aquila, A.%
, Ruti, P\BPBI M.%
\BCBL {}\ \BBA {} Speranza, A.%
\end{APACrefauthors}%
\unskip\
\newblock
\APACrefYearMonthDay{2007}{}{}.
\newblock
{\BBOQ}\APACrefatitle {Intercomparison of the northern hemisphere winter
  mid-latitude atmospheric variability of the IPCC models} {Intercomparison of
  the northern hemisphere winter mid-latitude atmospheric variability of the
  ipcc models}.{\BBCQ}
\newblock
\APACjournalVolNumPages{Climate Dynamics}{28}{7}{829--848}.
\newblock
\begin{APACrefURL} \url{https://doi.org/10.1007/s00382-006-0213-x}
  \end{APACrefURL}
\newblock
\begin{APACrefDOI} \doi{10.1007/s00382-006-0213-x} \end{APACrefDOI}
\PrintBackRefs{\CurrentBib}

\bibitem [\protect \citeauthoryear {%
Mehran%
, AghaKouchak%
\BCBL {}\ \BBA {} Phillips%
}{%
Mehran%
\ \protect \BOthers {.}}{%
{\protect \APACyear {2014}}%
}]{%
Mehran.2014}
\APACinsertmetastar {%
Mehran.2014}%
\begin{APACrefauthors}%
Mehran, A.%
, AghaKouchak, A.%
\BCBL {}\ \BBA {} Phillips, T\BPBI J.%
\end{APACrefauthors}%
\unskip\
\newblock
\APACrefYearMonthDay{2014}{}{}.
\newblock
{\BBOQ}\APACrefatitle {Evaluation of CMIP5 continental precipitation
  simulations relative to satellite-based gauge-adjusted observations}
  {Evaluation of cmip5 continental precipitation simulations relative to
  satellite-based gauge-adjusted observations}.{\BBCQ}
\newblock
\APACjournalVolNumPages{Journal of Geophysical Research:
  Atmospheres}{119}{4}{1695-–1707}.
\PrintBackRefs{\CurrentBib}

\bibitem [\protect \citeauthoryear {%
Monge%
}{%
Monge%
}{%
{\protect \APACyear {1781}}%
}]{%
Monge1781}
\APACinsertmetastar {%
Monge1781}%
\begin{APACrefauthors}%
Monge, G.%
\end{APACrefauthors}%
\unskip\
\newblock
\APACrefYearMonthDay{1781}{}{}.
\newblock
{\BBOQ}\APACrefatitle {{M\'emoire sur la th\'eorie des d\'eblais et des
  remblais}} {{M\'emoire sur la th\'eorie des d\'eblais et des
  remblais}}.{\BBCQ}
\newblock
\APACjournalVolNumPages{Histoire de l’Académie Royale des
  Sciences}{}{}{666-704}.
\PrintBackRefs{\CurrentBib}

\bibitem [\protect \citeauthoryear {%
Neelin%
, Langenbrunner%
, Meyerson%
, Hall%
\BCBL {}\ \BBA {} Berg%
}{%
Neelin%
\ \protect \BOthers {.}}{%
{\protect \APACyear {2013}}%
}]{%
Neelin.2013}
\APACinsertmetastar {%
Neelin.2013}%
\begin{APACrefauthors}%
Neelin, J\BPBI D.%
, Langenbrunner, B.%
, Meyerson, J\BPBI E.%
, Hall, A.%
\BCBL {}\ \BBA {} Berg, N.%
\end{APACrefauthors}%
\unskip\
\newblock
\APACrefYearMonthDay{2013}{}{}.
\newblock
{\BBOQ}\APACrefatitle {California winter precipitation change under global
  warming in the Coupled Model Intercomparison Project phase 5 ensemble}
  {California winter precipitation change under global warming in the coupled
  model intercomparison project phase 5 ensemble}.{\BBCQ}
\newblock
\APACjournalVolNumPages{Journal of Climate}{26}{17}{6238--6256}.
\PrintBackRefs{\CurrentBib}

\bibitem [\protect \citeauthoryear {%
Ning%
, Carli%
, Ebtehaj%
, Foufoula-Georgiou%
\BCBL {}\ \BBA {} Georgiou%
}{%
Ning%
\ \protect \BOthers {.}}{%
{\protect \APACyear {2014}}%
}]{%
Ning.ea.2014}
\APACinsertmetastar {%
Ning.ea.2014}%
\begin{APACrefauthors}%
Ning, L.%
, Carli, F\BPBI P.%
, Ebtehaj, A\BPBI M.%
, Foufoula-Georgiou, E.%
\BCBL {}\ \BBA {} Georgiou, T\BPBI T.%
\end{APACrefauthors}%
\unskip\
\newblock
\APACrefYearMonthDay{2014}{}{}.
\newblock
{\BBOQ}\APACrefatitle {Coping with model error in variational data assimilation
  using optimal mass transport} {Coping with model error in variational data
  assimilation using optimal mass transport}.{\BBCQ}
\newblock
\APACjournalVolNumPages{Water Resources Research}{50}{7}{5817-5830}.
\newblock
\begin{APACrefDOI} \doi{10.1002/2013WR014966} \end{APACrefDOI}
\PrintBackRefs{\CurrentBib}

\bibitem [\protect \citeauthoryear {%
Ott%
}{%
Ott%
}{%
{\protect \APACyear {1993}}%
}]{%
Ott1993}
\APACinsertmetastar {%
Ott1993}%
\begin{APACrefauthors}%
Ott, E.%
\end{APACrefauthors}%
\unskip\
\newblock
\APACrefYear{1993}.
\newblock
\APACrefbtitle {{Chaos in Dynamical Systems}} {{Chaos in Dynamical Systems}}.
\newblock
\APACaddressPublisher{Cambridge, UK}{Cambridge University Press}.
\PrintBackRefs{\CurrentBib}

\bibitem [\protect \citeauthoryear {%
Randall%
\ \protect \BOthers {.}}{%
Randall%
\ \protect \BOthers {.}}{%
{\protect \APACyear {2007}}%
}]{%
Randall2007}
\APACinsertmetastar {%
Randall2007}%
\begin{APACrefauthors}%
Randall, D.%
, Wood, R.%
, Bony, S.%
, Colman, R.%
, Fichefet, T.%
, Fyfe, J.%
\BDBL {}Taylor, K.%
\end{APACrefauthors}%
\unskip\
\newblock
\APACrefYearMonthDay{2007}{}{}.
\newblock
{\BBOQ}\APACrefatitle {{Climate Models and Their Evaluation}} {{Climate Models
  and Their Evaluation}}.{\BBCQ}
\newblock
\BIn{} S.~Solomon\ \BOthers {.}\ (\BEDS), \APACrefbtitle {{Climate Change 2007:
  The Physical Science Basis. Contribution of Working Group I to the Fourth
  Assessment Report of the Intergovernmental Panel on Climate Change}.}
  {{Climate Change 2007: The Physical Science Basis. Contribution of Working
  Group I to the Fourth Assessment Report of the Intergovernmental Panel on
  Climate Change}.}
\newblock
\APACaddressPublisher{Cambridge, UK and NY, USA}{Cambridge University Press}.
\PrintBackRefs{\CurrentBib}

\bibitem [\protect \citeauthoryear {%
Robin%
, Yiou%
\BCBL {}\ \BBA {} Naveau%
}{%
Robin%
\ \protect \BOthers {.}}{%
{\protect \APACyear {2017}}%
}]{%
Robin2017}
\APACinsertmetastar {%
Robin2017}%
\begin{APACrefauthors}%
Robin, Y.%
, Yiou, P.%
\BCBL {}\ \BBA {} Naveau, P.%
\end{APACrefauthors}%
\unskip\
\newblock
\APACrefYearMonthDay{2017}{}{}.
\newblock
{\BBOQ}\APACrefatitle {{Detecting changes in forced climate attractors with
  Wasserstein distance}} {{Detecting changes in forced climate attractors with
  Wasserstein distance}}.{\BBCQ}
\newblock
\APACjournalVolNumPages{Nonlinear Processes in Geophysics}{24}{}{393-405}.
\PrintBackRefs{\CurrentBib}

\bibitem [\protect \citeauthoryear {%
Santambrogio%
}{%
Santambrogio%
}{%
{\protect \APACyear {2015}}%
}]{%
Santambrogio2015}
\APACinsertmetastar {%
Santambrogio2015}%
\begin{APACrefauthors}%
Santambrogio, F.%
\end{APACrefauthors}%
\unskip\
\newblock
\APACrefYear{2015}.
\newblock
\APACrefbtitle {Optimal Transport for Applied Mathematicians: Calculus of
  Variations, PDEs, and Modeling} {Optimal transport for applied
  mathematicians: Calculus of variations, pdes, and modeling}.
\newblock
\APACaddressPublisher{}{Birkh\"auser, Basel}.
\PrintBackRefs{\CurrentBib}

\bibitem [\protect \citeauthoryear {%
Strogatz%
}{%
Strogatz%
}{%
{\protect \APACyear {2015}}%
}]{%
Strogatz2015}
\APACinsertmetastar {%
Strogatz2015}%
\begin{APACrefauthors}%
Strogatz, S\BPBI H.%
\end{APACrefauthors}%
\unskip\
\newblock
\APACrefYear{2015}.
\newblock
\APACrefbtitle {{Nonlinear Dynamics and Chaos: With Applications to Physics,
  Biology, Chemistry, and Engineering, 2nd Edition}} {{Nonlinear Dynamics and
  Chaos: With Applications to Physics, Biology, Chemistry, and Engineering, 2nd
  Edition}}.
\newblock
\APACaddressPublisher{Boulder, CO}{Westview Press}.
\PrintBackRefs{\CurrentBib}

\bibitem [\protect \citeauthoryear {%
Taylor%
, Stouffer%
\BCBL {}\ \BBA {} Meehl%
}{%
Taylor%
\ \protect \BOthers {.}}{%
{\protect \APACyear {2012}}%
}]{%
Taylor2012}
\APACinsertmetastar {%
Taylor2012}%
\begin{APACrefauthors}%
Taylor, K\BPBI E.%
, Stouffer, R\BPBI J.%
\BCBL {}\ \BBA {} Meehl, G\BPBI A.%
\end{APACrefauthors}%
\unskip\
\newblock
\APACrefYearMonthDay{2012}{}{}.
\newblock
{\BBOQ}\APACrefatitle {{An overview of CMIP5 and the experiment design}} {{An
  overview of CMIP5 and the experiment design}}.{\BBCQ}
\newblock
\APACjournalVolNumPages{Bulletin of the American Meteorological
  Society}{93}{}{485--498}.
\PrintBackRefs{\CurrentBib}

\bibitem [\protect \citeauthoryear {%
Tian%
\ \BBA {} Dong%
}{%
Tian%
\ \BBA {} Dong%
}{%
{\protect \APACyear {2020}}%
}]{%
Tian2020}
\APACinsertmetastar {%
Tian2020}%
\begin{APACrefauthors}%
Tian, B.%
\BCBT {}\ \BBA {} Dong, X.%
\end{APACrefauthors}%
\unskip\
\newblock
\APACrefYearMonthDay{2020}{}{}.
\newblock
{\BBOQ}\APACrefatitle {{The Double‐ITCZ Bias in CMIP3, CMIP5, and CMIP6
  Models Based on Annual Mean Precipitation}} {{The Double‐ITCZ Bias in
  CMIP3, CMIP5, and CMIP6 Models Based on Annual Mean Precipitation}}.{\BBCQ}
\newblock
\APACjournalVolNumPages{Geophysical Research Letters}{47}{}{}.
\PrintBackRefs{\CurrentBib}

\bibitem [\protect \citeauthoryear {%
Ulam%
}{%
Ulam%
}{%
{\protect \APACyear {1964}}%
}]{%
Ulam1964}
\APACinsertmetastar {%
Ulam1964}%
\begin{APACrefauthors}%
Ulam, S\BPBI M.%
\end{APACrefauthors}%
\unskip\
\newblock
\APACrefYear{1964}.
\newblock
\APACrefbtitle {{Problems in Modern Mathematics}} {{Problems in Modern
  Mathematics}}.
\newblock
\APACaddressPublisher{New York}{Science Edition Wiley}.
\PrintBackRefs{\CurrentBib}

\bibitem [\protect \citeauthoryear {%
Villani%
}{%
Villani%
}{%
{\protect \APACyear {2009}}%
}]{%
Villani2009}
\APACinsertmetastar {%
Villani2009}%
\begin{APACrefauthors}%
Villani, C.%
\end{APACrefauthors}%
\unskip\
\newblock
\APACrefYear{2009}.
\newblock
\APACrefbtitle {{Optimal Transport: Old and New}} {{Optimal Transport: Old and
  New}}.
\newblock
\APACaddressPublisher{Berlin Heidelberg, Germany}{Springer-Verlag}.
\PrintBackRefs{\CurrentBib}

\bibitem [\protect \citeauthoryear {%
Vissio%
}{%
Vissio%
}{%
{\protect \APACyear {2018}}%
}]{%
Vissio2018b}
\APACinsertmetastar {%
Vissio2018b}%
\begin{APACrefauthors}%
Vissio, G.%
\end{APACrefauthors}%
\unskip\
\newblock
\APACrefYearMonthDay{2018}{}{}.
\newblock
{\BBOQ}\APACrefatitle {{Statistical mechanical methods for parametrization in
  geophysical fluid dynamics}} {{Statistical mechanical methods for
  parametrization in geophysical fluid dynamics}}.{\BBCQ}
\newblock
\APACjournalVolNumPages{Reports on Earth System Science}{212}{}{}.
\PrintBackRefs{\CurrentBib}

\bibitem [\protect \citeauthoryear {%
Vissio%
\ \BBA {} Lucarini%
}{%
Vissio%
\ \BBA {} Lucarini%
}{%
{\protect \APACyear {2018}}%
}]{%
Vissio2018a}
\APACinsertmetastar {%
Vissio2018a}%
\begin{APACrefauthors}%
Vissio, G.%
\BCBT {}\ \BBA {} Lucarini, V.%
\end{APACrefauthors}%
\unskip\
\newblock
\APACrefYearMonthDay{2018}{}{}.
\newblock
{\BBOQ}\APACrefatitle {{Evaluating a stochastic parametrization for a
  fast–slow system using the Wasserstein distance}} {{Evaluating a stochastic
  parametrization for a fast–slow system using the Wasserstein
  distance}}.{\BBCQ}
\newblock
\APACjournalVolNumPages{Nonlinear Processes in Geophysics}{25}{}{413-427}.
\PrintBackRefs{\CurrentBib}

\end{thebibliography}

%
%
%
%
%

\end{document}